\documentclass[10pt,twocolumn,preprintnumbers,amsmath,amssymb,nofootinbib
,superscriptaddress]{revtex4}
\usepackage{color}

\newcommand{\be}{\begin{equation}}  
\newcommand{\ee}{\end{equation}}  
\newcommand{\bea}{\begin{eqnarray}}  
\newcommand{\eea}{\end{eqnarray}}  
\newcommand{\ba}{\begin{array}}  
\newcommand{\ea}{\end{array}}

\newcommand{\comment}[1]{}
\newcommand{\nn}{\nonumber}

\begin{document}
\preprint{FERMILAB-PUB-19-244-T}
\preprint{OUTP-19-07P}

\title {Scale independent $R^2$ inflation}

\author{Pedro G. Ferreira}
\email{pedro.ferreira@physics.ox.ac.uk}
\affiliation{Astrophysics, Department of Physics,
University of Oxford, Keble Road,
Oxford OX1 3RH\\}
\author{Christopher T. Hill}
\email{hill@fnal.gov}
\affiliation{Fermi National Accelerator Laboratory,
P.O. Box 500, Batavia, Illinois 60510, USA\\}
\author{Johannes Noller}
\email{johannes.noller@eth-its.ethz.ch}
\affiliation{Institute for Theoretical Studies, ETH Zurich, Clausiusstrasse 47, 8092 Zurich, Switzerland\\ }
\author{Graham G. Ross}
\email{g.ross1@physics.ox.ac.uk}
\affiliation{Rudolf Peierls Centre for Theoretical Physics, 
University of Oxford, 1 Keble Road,
Oxford OX1 3NP\\$ $}

\date{\today}

\begin{abstract}
Weyl (scale) invariant theories of scalars and gravity 
can generate all mass scales spontaneously. In this paper we study a particularly simple version -- scale invariant $R^2$ gravity -- and show that, during an inflationary period, it leads to fluctuations which, for a particular parameter choice,  are almost indistinguishable from normal $R^2$ inflation. Current observations place tight constraints on the primary coupling constant of this theory and predict a tensor to scalar ratio, $0.0033 > r > 0.0026$, which is testable with the next generation of cosmic microwave background experiments.
\end{abstract}

\maketitle

\section{Introduction}

Recently there has been a resurgence of interest in scale (Weyl) invariant theories as a possible solution to the hierarchy problem -- the need to keep the Brout, Englert, Higgs (BEH) boson \cite{Englert:1964et,Higgs:1964pj,Higgs:1964ia} light in the presence of gravity or large mass scales associated with physics beyond the Standard Model, such as Grand Unification. Such scale invariant theories have to generate all dimension-full scales through spontaneous breaking of the symmetry, including the Planck scale and the electroweak breaking scale associated with the BEH boson. To generate a period of inflation it is also necessary for the spontaneous scale symmetry breaking to give rise to the inflation scale and it has been shown that this is possible in the context of a model with two scalar fields non-minimally coupled to the Ricci scalar, $R$  \cite{ShapoZen,ShapoZen2,ShapoBlas,ShapoBell,Bezrukov:2012hx,GarciaBellido:2012zu,Rubio:2014wta,Trashorras:2016azl,Karananas:2016kyt,Casas:2017wjh,Casas:2018fum,Ferreira:2016vsc,Ferreira:2016wem,Ferreira:2016kxi,Ferreira:2018itt,Ferreira:2018qss,Ghilencea:2018thl,Ghilencea:2018dqd}. 
Such models can naturally lead to an acceptable period of inflation and, for the case in which one scalar is identified with the BEH boson, are similar to ``Higgs" inflation \cite{Bezrukov:2007ep}.

In this paper we are interested in a simpler possibility, generalising $R + R^2$ inflation to a scale independent form \cite{Maeda:1987xf,Wang:1990pr,Herrera:1995me,Rinaldi:2015uvu,Bamba:2015uxa,Tambalo:2016eqr,Karam:2018mft,Kubo:2018kho,Wetterich:2019qzx}. The Planck constraints on the inflationary observables are in remarkable agreement with the $R+R^2$ model of Starobinsky \cite{Starobinsky:1980te}, proposed in 1980 and with a spectrum of CMB perturbations as analysed by Mukhanov and Chibisov \cite{Mukhanov:1981xt} shortly afterwards. In particular the model predicts a spectrum with scalar perturbation index $n_s \sim 0.96$ and tensor to scalar ratio $r \sim 0.004$.

As an $R^2$ term is already scale invariant, to build a fully scale invariant form it is only necessary to add a single ``Jordan Brans Dicke" \cite{Jordan:1959,Brans:1961sx} scalar field that, after inflation, is the dominant source of spontaneous breaking of the scale symmetry and is responsible for generating the Planck mass. Since the $R^2$ term involves fourth order derivatives it implies the existence of an additional propagating scalar degree of freedom similar to the two scalar model mentioned above. In the locally symmetric case one scalar is eliminated, providing the longtitudinal degree of freedom of a massive''Weyl photon" \cite{Utiyama:1973nq,Smolin:1979uz,Nishino:2011zz} and this has recently been extended to include an $R^2$ term in \cite{Ghilencea:2019jux}.

In this paper we analyse the inflationary properties of both the global and local scale invariant $R + R^2$ models. Our analysis simplifies previous studies of scale invariant $R^2$ inflation \cite{Maeda:1987xf,Wang:1990pr,Herrera:1995me,Rinaldi:2015uvu,Bamba:2015uxa,Tambalo:2016eqr,Karam:2018mft,Kubo:2018kho,Wetterich:2019qzx}. through the observation that ``inertial" spontaneous Weyl symmetry breaking leads to the decoupling of the dilaton/Weyl photon and results in single field inflation with the second scalar being either a massless dilaton or providing the longitudinal component of a massive Weyl photon. In the global case the decoupling of the dilaton eliminates the usual ``fifth-force" constraints on Jordan Brans Dicke models \cite{Ferreira:2016kxi,Brax:2014baa}. Moreover the recently developed frame-independent approach \cite{Karamitsos:2017elm,Ferreira:2018qss} allows us to determine the inflationary parameters directly in the Jordan frame where the scale invariance is manifest. As a result we are able to give analytic forms for the inflationary parameters. We find that the inflationary predictions of both the global and local Weyl invariant models are the same but differ from the original $R+R^2$ Starobinsky model, except in a particular limit. 

In Section \ref{global} we construct the model that is globally invariant under scale (Weyl) symmetry. The origin of spontaneous ``inertial" Weyl symmetry breaking is discussed in Section \ref{inertial} where we show that this occurs independently of the scalar potential. In Section \ref{dilaton} we identify the dilaton, the Goldstone boson associated with the spontaneous breaking of the Weyl symmetry, and show that it decouples from the inflaton sector. Section \ref{local} analyses the locally Weyl symmetric model, and shows that it also undergoes inertial spontaneous symmetry breaking that generates a mass for the ``Weyl photon". The Weyl photon decouples from the inflaton sector which has the same form as found in the globally symmetric case. In Section \ref{inflation} it is shown that in a significant region of parameter space, slow-roll inflation occurs in both the globally and locally Weyl symmetric cases and gives an analytic form for the inflationary observables. The results are compared to the original Starobinsky model. Finally in Section V we present a summary and our conclusions.

\section{Globally Weyl invariant $R+R^2$ model.}\label{global}

\subsection{A Minimal Model}

The model consists of a real Jordan-Brans-Dicke scalar, $\phi$ non-minimally coupled to the Ricci scalar \cite{Jordan:1959,Brans:1961sx} plus the $R^2$ term introduced by Starobinsky \cite{Starobinsky:1980te}. 
The action is given by
\bea
\label{R2S}
S &=& \int d^4x \sqrt{-g} \left( {1\over 2}g^{\mu\nu}\partial_\mu\phi\partial_\nu\phi  \right.
        \nonumber \\
&& \qquad   \left. 
-\frac{\lambda}{4}\phi^4-{1\over 12}\alpha_1\phi^2R+ \frac{1}{6f_0^2} R^2 \right) 
\eea
This theory is the most general one relevant to the inflationary era that can be constructed from these fields that is invariant under the global Weyl (scale) transformation
\bea
\phi&\rightarrow&e^{-\epsilon} {\phi}\nonumber \\
g_{\mu\nu}&\rightarrow&e^{2\epsilon} { g}_{\mu\nu}
\label{WT}
\eea
Note that other, quadratic, terms could be included (involving the Ricci and Riemann tensors) but which are absent (or do not contribute to the overall dynamics) during the inflationary regime.

Since the $R^2$ term involves fourth order derivatives it contains an additional (scalar) degree of freedom.  To make this explicit it is convenient to reduce the fourth order derivatives to second order by introducing the
auxiliary field $\eta$\footnote{By redefining the fields on moving to the Einstein frame one obtains an explicit kinetic energy term for the new field $\eta$ with the canonical form for a scalar field coupled to gravity.}  \cite{Hindawi:1995an,Whitt:1984pd}  with the action now given by
\bea
\label{R3S}
   S &=&  \int d^4x \sqrt{-g} 
        \left( {1\over 2}g^{\mu\nu}\partial_\mu\phi\partial_\nu\phi -{\lambda\over 4}\phi^4
        \right. \nonumber \\
      &&\qquad \left.  -{1\over 12}\alpha_1\phi^2R
       - {1\over 12}\alpha_2\eta^2 R-{\xi\over 4}\eta^4 \right)  
   \eea
Here the equation of motion for $\eta$ gives 
\be
\eta^2=-{\alpha_2\over 6\xi}R
\label{eta0}
\ee
which, substituted in the eq.(\ref{R3S}), gives eq.(\ref{R2S}) for 
\be
\xi=6f_0^2({\alpha_2\over 12})^2
\label{B}
\ee
The action is very close to that found in the 2-scalar model analysed by several groups \cite{ShapoZen,ShapoZen2,ShapoBlas,ShapoBell,Ferreira:2016vsc,Ferreira:2016wem,Ferreira:2016kxi,Ferreira:2018itt,Ferreira:2018qss,Ghilencea:2018thl}; in the Jordan frame
the difference is the absence of a kinetic term for $\eta$.

\subsection{Inertial symmetry breaking}\label{inertial}

Under the Weyl transformation, $\eta\rightarrow e^{-\epsilon}\eta$ the Weyl current is given by
\be
K_\mu\equiv  \frac{1}{\sqrt{-g}}  {\delta S\over \delta \partial^\mu \epsilon}=(1-\alpha_1)\phi\partial_\mu\phi-\alpha_2\;\eta\partial_\mu\eta
\ee
It can be written as $K_\mu = \partial_\mu K$, where the kernel $K$ is given by
\be 
K={1\over 2}\left((1-\alpha_1)\phi^2-\alpha_2\eta^2\right)
\label{kernel}
\ee
and is covariantly conserved
\be
D^\mu K_\mu=0
\label{cons}
\ee
Consider now a patch of the universe which can be described as approximately spatially constant, but time dependent. These regions can be described by the Friedman- Robertson-Walker (FRW) metric corresponding to
$g_{\mu\nu}=[1,-a^2(t),-a^2(t),-a^2(t)]$.
 \ In terms of the kernel the conservation law of eq.(\ref{cons}) becomes
 \be
 \ddot K+3H\dot K=0
 \ee
 where $H={\dot a}/a$, 
giving
\be
K(t)=c_1+c_2\int_{t_0}^t{dt'\over a^3(t')}
\ee
 where $c_1$ and $c_2$ are constants. Thus in an expanding universe $K(t)$ will evolve to a constant value, $\bar{K} = {K}(t\rightarrow\infty)$ that, from eq.(\ref{kernel}) implies the scalar fields acquire constant vacuum expectation values, spontaneously breaking Weyl symmetry \cite{Ferreira:2018itt}. 
 
 Note that this inertial Weyl symmetry breaking does not rely on a scalar potential, the value of the kernel being determined by the chaotic initial conditions. Note also that a constant kernel implies a relation between the scalar fields which is the reason the model results in single field inflation.

\subsection{Dilaton Decoupling}\label{dilaton}

To identify the Goldstone mode associated with the spontaneous breaking of the global Weyl symmetry we change variables to 
\bea
\phi&=&e^{-\sigma(x)/f} \hat\phi\nonumber \\
\eta&=&e^{-\sigma(x)/f} \hat\eta\nonumber \\
g_{\mu\nu}&=&e^{2\sigma(x)/f}  \hat g_{\mu\nu}
\label{variable}
\eea
 where $f$ has dimensions of mass. Note that $\hat\phi,\;\hat\eta$ and $\hat g_{\mu\nu}$ are invariant under the global Weyl symmetry and only $\sigma$ transforms, $\sigma\rightarrow\sigma +c$ where $c=\epsilon f$. In terms of the new metric
 \be
R(g)=e^{2\sigma/f}\left( R(\hat g)+6\hat D^2\sigma/f+6(\hat D\sigma/f)^2\right)
\ee
giving
\bea
S &=&  \int d^4x \sqrt{-g}
\left( - \frac{1}{12}({\alpha _1}{{\hat \phi }^2} + {\alpha _2}{{\hat \eta }^2})\hat R
+ \frac{1}{2}{\partial _\mu }\hat \phi {\partial ^\mu }\hat \phi 
\right.
\nonumber \\
&& \left.
 + \frac{1}{{f^2}}\hat K{\partial _\mu }\sigma {\partial ^\mu }\sigma  + \frac{1}{f}{\partial _\mu }\sigma {\partial ^\mu }\hat K - \frac{\lambda }{4}{{\hat \phi }^4} - \frac{\xi }{4}{{\hat \eta }^4}\right.\nonumber\\
&&\left. +\lambda_LC(\hat\phi,\hat\eta)\right)
\eea
where we have added a Lagrange multiplier, $\lambda_L$, and the constraint 
\be
C(\hat\phi,\hat\eta)=\bar K -\frac{1}{2}\left( {(1 - {\alpha _1}){{\hat \phi }^2} - {\alpha _2}{{\hat \eta }^2}} \right)
\label{khat}
\ee
For the constant $\bar K$ this gives the ``Einstein frame'' form of the theory
\bea
S &=& \int d^4x \sqrt { -\hat g} \left( { - \frac{1}{{12}}({\alpha _1}{{\hat \phi }^2} + {\alpha _2}{{\hat \eta }^2})\hat R} + \frac{1}{2}{\partial _\mu }\hat \phi {\partial ^\mu }\hat \phi \right. \nonumber \\
&& \left.
 - \frac{\lambda }{4}{{\hat \phi }^4} - \frac{\xi }{4}{{\hat \eta }^4} + \frac{1}{2}{\partial _\mu }\sigma {\partial ^\mu }\sigma +\lambda_LC(\hat\phi,\hat\eta) \right)
\label{EFaction}
\eea
where the dilaton is canonically normalized with the choice $f^2=2\bar {K} $.

The field $\sigma$ is the massless dilaton. We see that it decouples from the hatted scalar fields. The same is true if one adds vector and fermion fields to the theory \cite{Ferreira:2018itt}. As a result one avoids the severe astrophysical constraints on the fifth force normally associated with the dilaton.

Moreover, note that the dilaton now yields the Weyl current in the Einstein frame. 
We see that under the Weyl
transformation $\delta\sigma/f = \delta\epsilon$, where all hatted variables are invariant,
the Weyl current is now:
\bea
K_\mu\equiv  \frac{1}{\sqrt{-g}} \frac{\delta S}{ \delta \partial^\mu \epsilon} = f\partial_\mu\sigma
\eea
which is the familiar form of the current of a Nambu-Goldstone boson.
Inertial symmetry breaking can be understood in the ``Einstein frame'' 
as a chaotic initial $K(x)$ parameterized by an arbitrary constant $\bar{K}$ and an initially
chaotic dilaton field $\sigma(x)$, as $K=e^{2\sigma(x)/f}\bar{K}$, with $f^2=2\bar{K} $.
This dilaton form of the current,  $f\partial_\mu\sigma $, is covariantly
conserved, equivalent to the massless dilaton equation of motion, $D^2 \sigma =0$. This
implies that the dilaton redshifts to a constant, yielding the spontaneously broken
scale symmetry.

\section{ LOCALLY WEYL INVARIANT $R + R^2$ MODEL}\label{local}
For the case of local Weyl transformations, $\epsilon=\epsilon(x)$ in eq.(\ref{WT}), the action, eq.(\ref{R2S}), is no longer invariant. To correct this it is necessary to introduce a vector field, $A_\mu$, the ``Weyl photon" that transforms as
\be
A_\mu\rightarrow A_\mu-\partial_\mu \epsilon (x)
\ee
With this one can define a Weyl derivative acting on scalar fields, $\phi$, by 
${\tilde D_\mu }\phi  \equiv ({\partial _\mu } - {A_\mu })\phi $ that transforms covariantly
\be
\tilde D_\mu\phi\rightarrow e^{-\epsilon(x)}\tilde D_\mu \phi
\ee
and replaces the Riemannian derivative when constructing a locally invariant version of eq.(\ref{R2S}). Similarly it is necessary to replace the Riemannian Ricci scalar, $R$, with its Weyl geometric version, $\tilde R$ defined by
\be
\tilde R = R - 6{D_\mu }{A^\mu } - 6{A_\mu }{A_\mu }
\label{tildeR}
\ee
 where $D_\mu$ is the Riemannian covariant derivative. Now $\tilde R$ transforms covariantly under local Weyl transformations
 \be
\tilde R\rightarrow e^{-2\epsilon(x)}\tilde R
\ee
 Using these covariant forms the locally invariant version of eq.(\ref{R2S}) is \cite{Ghilencea:2019jux,Ghilencea:2018dqd} 
\bea
{S_L} &=& 
\int {d^4}x\sqrt {- g} \left( \frac{1}{2}{g^{\mu \nu }}{{\tilde D}_\mu }\phi {{\tilde D}_\nu }\phi  - \frac{\lambda }{4}{\phi ^4} 
\right. \nonumber \\
&& \left.
- \frac{1}{{12}}{\alpha _1}{\phi ^2}\tilde R + \frac{1}{{6f_0^2}}{{\tilde R}^2} - \frac{1}{{4{e^2}}}{F^{\mu \nu }}{F_{\mu \nu }}\right)
\eea
where we have added to the Lagrangian the Weyl invariant Weyl photon kinetic term, $-{1\over 4e^2}F^{\mu\nu}F_{\mu\nu}$ where $F_{\mu\nu}=\partial_\mu A_\nu-\partial_\nu A_\mu$. Introducing an auxiliary field as above this can be rewritten as
\bea
S_L &\equiv & \int d^4 x \sqrt { - g} 
\left(\frac{1}{2}g^{\mu \nu }{{\tilde D}_\mu }\phi {{\tilde D}_\nu }\phi  - \frac{\lambda }{4}{\phi ^4} 
\right. \nonumber \\
& & \!\!\!\!\!\!\!\!\!   \left.
- \frac{1}{12}({\alpha _1}{\phi ^2} + {\alpha _2}{\eta ^2})\tilde R - \frac{\xi }{4}{\eta ^4} - \frac{1}{4{e^2}} {F^{\mu \nu }}{F_{\mu \nu }} \right)
\eea
Substituting $\tilde R$, using eq.(\ref{tildeR}), integrating by parts,
and rescaling  $A_\mu \rightarrow eA_\mu$ gives
\bea
\label{SL0}
{S_L} & =& \int {d^4}x\sqrt { - g} \left( {\frac{1}{2}{\partial _\mu }\phi {\partial ^\mu }\phi  - \frac{1}{{4}}{F^{\mu \nu }}{F_{\mu \nu }} - \frac{1}{2}e{A_\mu }{K^\mu } }\right.
\nonumber \\
& & \!\!\!\!\!\!\!\!\!  \!\!\!\!\! \!\!\!\!\!   \left. 
 + \frac{1}{2}{e^2}K {A_\mu }{A^\mu } - \frac{1}{{12}}({\alpha _1}{\phi ^2} + {\alpha _2}{\eta ^2})R - \frac{\lambda }{4}{\phi ^4} - \frac{\xi }{4}{\eta ^4} \right) 
\eea
where
\be
{K_\mu } = {\partial _\mu }K,\qquad K = \frac{1}{2}\left((1 + {\alpha _1}){\phi ^2} 
+ {\alpha _2}{\eta ^2}\right)
\ee
Note that the variation of the action or eq.(\ref{SL0}) with respect to $\partial_\mu\epsilon(x)$,
with  $\delta eA_\mu= \partial_\mu \epsilon (x)$,
now yields the vanishing identity:
\bea
\frac{1}{{\sqrt {- g} }}\frac{{\delta S_L}}{{\delta {\partial^\mu\epsilon }}}
= K_\mu -  {\partial _\mu }K=0
\eea
In any local gauge theory the variation of the full action
with respect to the local gauge angle always
produces zero, since this is the very definition of the symmetry. 

The conserved Weyl current in the global case
is now replaced by the Weyl photon equation of motion
\be
  \frac{1}{{\sqrt { - g} }}\frac{{\delta S_L}}{{\delta {A^\mu }}} 
  = D_\nu F^{\mu\nu} - 2{A_\mu }{e^2}K- e{K_\mu }=0
\label{weylcurrent}
\ee
  Changing variables as in eq.(\ref{variable}), for constant $\bar K$, the action becomes
\bea
S_L &=& \int \sqrt g \left( { - \frac{1}{{12}}({\alpha _1}{{\hat \phi }^2} + {\alpha _2}{{\hat \eta }^2})\hat R + \frac{1}{2}{\partial _\mu }\hat \phi {\partial ^\mu }\hat \phi  
}- \frac{\lambda }{4}{{\hat \phi }^4}\right. \nonumber \\
&& \left. - \frac{\xi }{4}{{\hat \eta }^4} - \frac{1}{{4}}{F^{\mu \nu }}{F_{\mu \nu }} + \frac{1}{2}e^2\bar K\;{B_\mu }{B^\mu+\lambda_L C } \right)
\eea
where $B_\mu=A_\mu-{e\over f}\partial_\mu\sigma$.  

The spontaneous breaking of the Weyl symmetry has generated a mass for the Weyl photon,
$M^2=e^2\bar{K}$, with the would-be Goldstone mode, $\sigma$, providing its longitudinal component. As for the dilaton in the global case the Weyl photon decouples from the hatted fields.  In the present local case the dilaton, $\sigma$
and the gauge field, $A_\mu$,  are not separately physical. They are replaced by the combined
(Stueckelberg) field $B_\mu$ which is physical and invariant under local Weyl transformations.

Note that the conserved Weyl current is now given by the divergence
of the field $B_\mu$ itself, which satisfies
the Weyl Maxwell equation:
\be
D_\nu F^{\mu\nu} - M^2{B_\mu }=0  \qquad M^2=e^2\bar{K}
\ee
A massive spin-1 field necessarily obeys the Lorentz gauge condition
$D_\mu B^\mu =0$ owing to the anti-symmetry of $F_{\mu\nu}$ (note that a massive
spin-1 field has mass term $-M^2B_\mu$, contrary to a scalar $+M^2\phi$).
Hence an arbitrary initial physical $B_\mu$ will redshift to zero, yielding the
spontaneously broken scale symmetry vacuum, with $B_\mu=0$, and  which is the analogue
of the redshifting of $K_\mu\rightarrow 0$ and  $K_\mu\rightarrow \bar{K}$ in the global case.
The vacuum is analogous to a superconductor, where the massive $B_\mu$ has the solution
$B_\mu=0$, which is the analogue of the London equation for a superconductor,
$e\vec{A}-\vec{j}=0$.

\section{INFLATION}\label{inflation}

It is now straightforward to analyse the possibility that there is a period of inflation in both the global and local cases. The decoupled fields, the dilaton in the global case or the Weyl photon in the local case, play no role in the inflationary era. In their absence the action for the hatted fields is the same for both cases and so their inflationary structure is the same.

In the Jordan frame the Klein Gordon equations resulting from eq.(27) are given by
\bea
{D^2}\hat \phi  + \lambda {\hat \phi ^3} + \frac{1}{6}{\alpha _1}\hat \phi R &=& 0\nonumber\\
 \xi {\hat \eta ^3} + \frac{1}{6}{\alpha _2}\hat \eta R &=& 0
\eea
Thus
\be
{\alpha _2}\frac{{{D^2}\hat \phi }}{{\hat \phi }} =  - {\alpha _2}\lambda {\hat \phi ^2} + {\alpha _1}\xi {\hat \eta ^2}
\label{roll1}
\ee
Note that there is an infrared stable fixed point that determines the ratio of the field vacuum
expectation values (VEVs) given by 
\be
{\hat\eta^2\over\hat\phi^2}={\alpha_2\lambda\over\alpha_1\xi}
\ee
that is approached after any period of inflation \cite{Ferreira:2016vsc} .

Taking the slow-roll limit with the ``inflation derivative" (i.e. in terms of the number of e-foldings) $ {D^2}\hat \phi  \to 3H\dot \hat \phi  = 3{H^2}{\partial _N}\hat \phi $ where $ N = \ln a(t) $ gives
\be
{3\over 2}H^2\alpha_2{\partial_N\hat\phi^2\over\hat\phi^2}=-\alpha_2\lambda\hat\phi^2+\alpha_1\xi\hat\eta^2
\ee
$H^2$ may be eliminated using the (00) Einstein equation in the slow-roll limit
\be
G_{00}=3H^2\approx-{3\over 2}g_{00}{\lambda\hat\phi^4+\xi\hat\eta^4\over\alpha_1\hat\phi^2+\alpha_2\hat\eta^2}
\ee
From eq.(\ref{khat}) one sees that the constraint that the Kernel should be constant corresponds to the fields being constrained to move on an ellipse. Without loss of generality we can choose the ellipse $\bar K = 1$ and map the first quadrant of the ellipse into the variables 
\be
x=(1-\alpha_1)\hat\phi^2,\;\;y=-\alpha_2\hat\eta^2,
\label{roll2}
\ee
so that the ellipse becomes the line segment $x + y = 1$. Using eqs(\ref{roll1}-\ref{roll2}) we find
  \be
  \partial_Nx={S(x)\over W(x)}x(1-x)(x-x_0)
  \label{xevolution}
  \ee
    where
    \be
    S(x)={1\over 3}\times{-\alpha_2 \lambda+\alpha_1(1-\alpha_1)\xi\over (1-\alpha_1)^2\alpha_2^2} \times{1-x-\alpha_1\over 1-x}
    \ee 
  
    \be
    x_0={\xi(1-\alpha_1)\over \xi(1-\alpha_1)-\alpha_2^2\lambda}
    \ee
    and 
    \be
    W(x)={\lambda x^2\over 4(1-\alpha_1)^2}+{\xi(1-x)^2\over 4\alpha_2^2}   
     \ee
    There are various parameter choices that can give slow roll inflation. For the case in which one coupling dominates, $\partial_Nx$ is independent of the coupling. For the case in which $\xi$ dominates,  $\partial_Nx\propto {\alpha_1\over 1-\alpha_1}$ and slow roll can occur for $\alpha_1$ small. For the case in which $\lambda$ dominates  $\partial_Nx\propto {1-x-\alpha_1\over \alpha_2}$ and slow roll can occur for $\alpha_2$ large. 
Here we examine the former case with $\lambda\ll\xi$ and $\alpha_1$ small which makes it easy to compare to the 2-scalar case that assumed the same limit \cite{Ferreira:2018qss}. Keeping only $\xi$ non-zero we have
\be
{\partial _N}x=  - \frac{4}{3}{\alpha _1}x\ee  
  \be
     H^2\approx {\xi y^2\over 2\alpha_2^2(y-\alpha_1 x)}    
     \label{hubble}
      \ee    
 and $x(t)$ will roll from an initial $x(0)$ toward $x(t_E) \equiv  x_E \approx 1$. Here $t_E$ is the end of inflation given by the inflation parameter, $\epsilon$, approaching unity: $\epsilon=-{1\over 2}{d\ln H^2\over dN}\approx 1$ corresponding to $y\approx -{2\over 3}\alpha_1$ or, equivalently, $\alpha_1\hat\phi^2\approx\alpha_2\hat\eta^2$. 
Thus
  \be \label{phisol}
     x=x_Ee^{-\nu N_J},\;\;y=1-x_Ee^{-\nu N_J},\;\;\nu=-{4\over 3}\alpha_1
     \ee
where $N_J$ is the number of e-foldings until the end of inflation in the Jordan frame. This is the same form as was found for the 2-scalar model.

For clarity, the discussion to date has been in the Jordan frame. To determine the slow roll parameters we use the frame independent formalism introduced in \cite{Karamitsos:2017elm} and as applied in  \cite{Ferreira:2018qss}. Denoting $\theta_1=\hat\phi,\;\theta_2=\hat\eta$, the frame-invariant field-space metric $G_{AB}$ and potential $U$ are given by
\be \label{EffDef}
{G_{AB}} = \frac{{{\delta _{11}}}}{f} + \frac{3}{2}\frac{{{f_{,A}}{f_{,B}}}}{{{f^2}}},\;\;\;U = \frac{{\lambda \;\theta _1^4 + \xi \;\theta _2^4}}{{4{f^2}}},
\ee
{where the key difference to the two-field model considered in \cite{Ferreira:2018qss} is the occurrence of $\delta_{11}$ (instead of $\delta_{AB}$) and where}
\be \label{fdef}
f \equiv M_P^2 \equiv  - \frac{1}{6}\sum\limits_{A = 1}^2 {{\alpha _A}} \theta _A^2.
\ee

Defining the scalar (tensor) spectral index, $n_S$ ($n_T$), the running of the scalar (tensor) spectral index, $\alpha_S$ ($\alpha_T$) and the tensor to scalar ratio, $r$, we have that the frame independent analysis proceeds as before, giving for the inflation observables
\bea
n_S &=& 1 + \frac{4\alpha_1 (e^{-\nu N_J}+1)}{3(1 - e^{-\nu N_J})} +  {\cal O}(\alpha_1^2), \nonumber \\
r &=& \frac{64 \alpha_1^2  e^{-\nu N_J}}{3  (e^{-\nu N_J} - 1)^2} + {\cal O}(\alpha_1^3), \nonumber \\
\alpha_S &=&- \frac{32 \alpha_1^2 e^{-\nu N_J}}{9 (e^{-\nu N_J}-1)^2} + {\cal O}(\alpha_1^3), \nonumber \\
n_T &= &-  \frac{8 \alpha_1^2  e^{-\nu N_J}}{3  (e^{-\nu N_J} - 1)^2}  + {\cal O}(\alpha_1^3) , \nonumber \\
\alpha_T &=& - \frac{32 \alpha_1^3  e^{-\nu N_J} (1 + e^{-\nu N_J})}{9 (e^{-\nu N_J} - 1)^3}  + {\cal O}(\alpha_1^3), 
\label{obsExp}
\eea
 where the relation between the number of e-foldings in the Einstein and Jordan frames is given by
 \begin{align} \label{Nsol2}
N_E &= N_J + \frac{1}{2} \ln\left(\frac{2\alpha_1 }{(e^{-\nu N_J} - 1 )}\right)
+ {\cal O}(\alpha_1) \\
&= {N_J + \frac{1}{2} \ln \left(\frac{3}{2N_J}\right) + {\cal O}(\alpha_1)}, \nn
\end{align}
{where we have expanded the exponential for small $\alpha_1$ in the second line and assume $N_J \gg 1$. For $N_J = 60$, we therefore obtain $N_E \sim 58$.}
{One feature that is immediately apparent is the absence of $\alpha_2$ in \eqref{obsExp}. Indeed we have explicitly checked up to $10$th order in $\alpha_1$ that no such dependence is present. In Appendix \ref{app:limit} we show why this is the case, but here we simply emphasise that despite the appearance of four parameters $(\alpha_1,\alpha_2,\lambda,\xi)$, in the $\xi \gg \lambda$ limit we are considering here only one single parameter controls all of the above observables. Working in this limit then also enforces a tight observational bound on $\alpha_1$: Current constraints on $n_S$ (Planck 2018 \cite{Aghanim:2018eyx,Akrami:2018odb} finds $n_S=0.9649 \pm 0.0042$) enforce $|\alpha_1| < 0.019$.\footnote{For the model considered here there is also an upper bound on $n_S$, namely $n_S \lesssim 1-2/N_J\simeq 0.9663$ (where we are assuming $N_J\simeq60$), which is saturated as $\alpha_1\rightarrow 0$.} The tensor-to-scalar ratio $r$, on the other hand, is always comfortably within current observational bounds ($r \lesssim 0.064$ -- see \cite{Aghanim:2018eyx,Akrami:2018odb}). In fact, the precise value for $r$ can be accurately predicted for our model: $0.0033 > r > 0.0026$, where the lower bound is a consequence of the $n_S$ induced bound on $\alpha_1$ and the upper bound is saturated for $\alpha_1 \sim 0$ (and we again assume $N_J\simeq60$))  

The remaining observables in \eqref{obsExp} satisfy $a_S \sim -2/N_J^2 \sim -5 \cdot 10^{-4}$, $a_T \sim - 10^{-5}$ and $n_T \sim - 4 \cdot 10^{-4}$, for all allowed choices of $\alpha_1$. Also, to reproduce the observed magnitude of the fluctuations it is neccessary to choose ${\xi\over\alpha_2^2}=O(10^{-10})$.
}

{
The above results may be compared to the original Starobinsky model. Focusing on $n_S$ and $r$, there one finds
\be
n_s-1\approx-{2\over N_E},\;\;r\approx{12\over N_E^2}.
\ee
Expanding the observables in \eqref{obsExp} in the limit $\alpha_1\rightarrow 0$ for general $N_J$ (assumed to satisfy $N_J \gg 1$), the scale invariant model gives
\be
n_s-1\approx-{2\over N_J},\;\;r\approx{12\over N_J^2}
\ee
{at leading order in $N_J^{-1}$}, so that the predictions are very close to the original Starobinsky model (recall that $N_J$ and $N_E$ only differ by $\sim 3\%$ in the scale invariant model investigated here, assuming close to 60 e-folds of inflation).
}

{
Finally a note on observables beyond the 2-point function and on isocurvature perturbations. The essential features here are identical to the ones of the scale-independent two field models considered in \cite{Ferreira:2018qss}: The local non-Gaussian parameter $f_{NL}$ satisfies \cite{Elliston:2012ab,Byrnes:2012sc} 
\begin{align}
f_{NL}^{\rm local} &\approx \frac{5}{6}\frac{N^{,A}N^{,B}(\nabla_A\nabla_BN)}{(N_{,C}N^{,C})^2},
\end{align}
where $N$ is the frame-covariant number of e-folds ($N = N_E$) \cite{Karamitsos:2017elm}. Evaluating this expression and expanding for small $\alpha_1$ we find
\begin{equation}
f_{NL}^{\rm local} \approx \frac{5 \alpha_1 (e^{-\nu N_J}+1)}{9 (e^{-\nu N_J}-1)} + {\cal O}(\alpha_1^2) \sim 1.5 \times 10^{-2},
\end{equation}
so no sizeable non-Gaussian signature. Indeed this is to be expected from the existence of a physically equivalent single scalar theory with a canonical kinetic term (see appendix \ref{app:single}). The effective single field nature then mandates suppressed local non-Gaussianity \cite{Maldacena:2002vr,Acquaviva:2002ud}, while the canonical nature of the kinetic term ensures the absence of sizeable equilateral non-Gaussianity (because, for general single-field models, $f_{NL}^{\rm equil} \lesssim 1/c_s^2$).
Significant isocurvature perturbations are also absent, since the acceleration vector between paths in the geodesic flow
\begin{eqnarray}
\omega^A&=&X^B\nabla_B\left[\frac{X^A}{\sqrt{X_C X^C}}\right] = 0
\label{isoparams1}
\end{eqnarray}
for the models considered here (up to second order in slow-roll), where we have defined $X_A \equiv (\ln U)_{,A}$. When $\omega^A = 0$, the transfer function that converts curvature perturbations into isocurvature perturbations vanishes \cite{Karamitsos:2017elm,Ferreira:2018qss}, so no sizeable isocurvature perturbations are generated for these models. 
}

\section{ SUMMARY AND CONCLUSIONS}
We have studied the scale free version of $R^2$ inflation, setting it within the framework of the scale-invariant theories extensively studied in \cite{ShapoZen,ShapoZen2,ShapoBlas,ShapoBell,Bezrukov:2012hx,GarciaBellido:2012zu,Rubio:2014wta,Trashorras:2016azl,Karananas:2016kyt,Casas:2017wjh,Casas:2018fum,Ferreira:2016vsc,Ferreira:2016wem,Ferreira:2016kxi,Ferreira:2018itt,Ferreira:2018qss,Ghilencea:2018thl,Ghilencea:2019jux,Ghilencea:2018dqd}.  It can be shown to correspond to the two field model but without a canonical kinetic term for one of the scalar fields in the Jordan frame (for a more careful analysis of this limit, see Appendix \ref{app:limit}). 
As a consequence, we have shown that scale-free $R^2$ inflation is endowed with inertial symmetry breaking of Weyl invariance and that the techniques that were developed in \cite{Ferreira:2016vsc,Ferreira:2016wem} can be applied seamlessly to this scenario and used to derive predictions for the inflationary observables. The model we have considered here is, to some extent, the simplest non-minimally coupled theory which is scale invariant; it falls into the class of models considered in \cite{KS2014} which are favoured by the Planck constraints.

The effect of radiative corrections on the inflationary era have been briefly discussed in the context of the two scalar model \cite{Ferreira:2016wem,Ferreira:2016kxi} 
 and the general structure applies to the model considered here. In particular, if the scale invariance is preserved by quantum corrections by a suitable choice of the regularisation procedure that does not introduce an extrinsic mass scale, inertial symmetry breaking still occurs and the dilaton still decouples. There will still be a period of slow roll inflation but the form of the constraint between the hatted fields will change and the ellipse on which they move will be distorted \cite{Ferreira:2016kxi} giving a small change in the inflationary predictions provided the couplings are in the perturbative  range.

We have shown that the inflationary observables -- $n_S$, $r$, $\alpha_S$, $n_T$ and $\alpha_T$ -- are uniquely determined in terms of the small parameter, $\alpha_1$, which controls the non-minimal coupling of the extra scalar field. As a result, and with current constraints on the scalar spectral index, $n_S$, we can place tight bounds, $|\alpha_1|<0.019$, which in turn leads to a very clear prediction for the tensor to scalar ratio: $0.0033 > r > 0.0026$. This small range of values is within reach of the constraining power of future experiments: while the Simons Observatory \cite{SimonsObs} has a forecast sensitivity of $\sigma(r) = 0.003$, the LITEBIRD Mission  will push the sensitivity to $\sigma(r) = 0.001$, while the S4 CMB consortium will aim for $\sigma(r) = 0.0005$ \cite{rforecasts}. This means that this model offers a characteristic and precise prediction that will be experimentally testable within the next decade.
\\

\textit{Acknowledgements ---} PGF is supported by the ERC and the Beecroft Trust. JN acknowledges support from Dr.\ Max R\"ossler, the Walter Haefner Foundation and the ETH Zurich Foundation. 
CTH acknowledges Fermilab, operated by Fermi Research Alliance,  
LLC under Contract No. DE-AC02-07CH11359 with the United States Department of
Energy.

\appendix

\section{Scale independent $R^2$ inflation as a limit of a bi-scalar theory}
\label{app:limit}

{Here we formally relate the results obtained for the scale-independent $R^2$ models considered throughout this paper to the scale-independent two field models considered in \cite{Ferreira:2018qss}. Recall the auxiliary field representation \eqref{R3S} 
\begin{align}
S &=  \int d^4x \sqrt{-g} \Big(  -\frac{1}{12}(\alpha_1\hat\phi^2 +\alpha_2\hat\eta^2) R \nonumber \\ &+ \frac{1}{2}\partial_\mu\hat\phi\partial^\mu\hat\phi -{\lambda\over 4}\hat\phi^4  -{\xi\over 4}\hat\eta^4 \Big) 
\label{R3Sv2} 
\end{align}
and add an explicit kinetic term for the $\eta$ field in the Jordan frame. One then finds
\begin{align}
S &=  \int d^4x \sqrt{-g} \Big(  -\frac{1}{12}(\alpha_1\hat\phi^2 +\alpha_2\hat\eta^2) R \nonumber \\ &+ \frac{1}{2}\partial_\mu\hat\phi\partial^\mu\hat\phi + \frac{1}{2}\kappa^2\partial_\mu\hat\eta\partial^\mu\hat\eta -{\lambda\over 4}\hat\phi^4  -{\xi\over 4}\hat\eta^4 \Big),  
\label{R4S}
\end{align}
where we have introduced a constant parameter $\kappa$. We may now canonically normalise the $\hat\eta$ field by sending $\hat\eta \to \hat\eta/\kappa$ and then find
\begin{align}
S &=  \int d^4x \sqrt{-g} \Big(  -\frac{1}{12}(\alpha_1\hat\phi^2 +\hat\alpha_2\hat\eta^2) R \nonumber \\ &+ \frac{1}{2}\partial_\mu\hat\phi\partial^\mu\hat\phi + \frac{1}{2}\partial_\mu\hat\eta\partial^\mu\hat\eta -{\lambda\over 4}\hat\phi^4  -\frac{\hat\xi}{4}\hat\eta^4 \Big),
\label{R4Sv2}  
\end{align}
which is precisely of the type considered in \cite{Ferreira:2018qss} and where 
\begin{align}
\hat\alpha_2 &\equiv \frac{\alpha_2}{\kappa^2}, &\hat\xi &\equiv \frac{\xi}{4 \kappa^4}
\end{align}
The model \eqref{R3Sv2}, i.e. the $\kappa \to 0$ limit of \eqref{R4S}, is therefore analogous to sending $\hat\alpha_2$ and $\hat\xi$ to $\pm \infty$ in \eqref{R4Sv2} (with the appropriate scalings determined by their $\kappa$-dependence). Observables computed for \eqref{R3Sv2} and \eqref{R4Sv2} can therefore be related by taking appropriate scaling limits of the $\hat\alpha$ and $\hat\xi$ parameters and indeed, as expected from this argument, the expressions \eqref{obsExp} are precisely the $|\alpha_2| \to \infty$ limit of the analogous expressions obtained in \cite{Ferreira:2018qss}.} Note that there remains an $\alpha_2$ dependence in determining the magnitude of the fluctuations as it is proportional to ${\xi\over\alpha_2^2}\equiv {\hat\xi\over \hat\alpha_2^2}$.

\section{Effective single scalar theory}
\label{app:single}

Using eq(\ref{khat}) with $\bar K$ constant we can express the field $\hat \eta$ in terms of $\hat\phi$ and describe the hatted field action that drives inflation in the scale independent $R^2$ theory as a (single-)scalar-tensor theory:%
\begin{eqnarray}
S&=&\int d^4 x\sqrt{-g}\Big[\frac{\hat f({\hat\phi})}{2} \hat R+\frac{1}{2}\partial_\mu\hat\phi\partial^\mu \hat\phi - \hat W({\hat\phi})\Big]
\label{Ssingle}
\end{eqnarray}
As discussed in section \ref{inflation}, the form of \eqref{Ssingle} immediately ensures the absence of a sizeable non-Gaussian signature for this model and hence for \eqref{EFaction}. 

The functions $\hat f(\hat\phi)$ and $\hat W(\hat\phi)$ are given by
\be
\hat f = \frac{1}{6} (2\bar K -  \hat\phi^2), \;\;
\hat W = \frac{\xi}{\alpha_2^2} \bigl(2\bar K + (\alpha_1 - 1) \hat\phi^2\bigr)^2
\label{modfunc}
\ee
where we may now define the frame-invariant potential $\hat U \equiv \hat W/\hat f^2$ and can also evaluate the frame invariant field space metric $\hat G$, c.f. \eqref{EffDef}, which here is a simple scalar function (since the scalar field space is one dimensional now). We find
\begin{equation}
\hat G = \frac{12 \bar K}{(\hat\phi^2 - 2\bar K)^2},
\end{equation}
where the frame-invariant inverse `metric' satisfies $\hat G^{-1} = 1/\hat G$. Using the same frame independent analysis as above (and as explained in detail in \cite{Karamitsos:2017elm,Ferreira:2018qss}), we can now compute the inflationary observables $n_S$, $r$, $\alpha_S$, $n_T$, $\alpha_T$ and indeed find identical predictions for the effective single field model \eqref{Ssingle} when compared with the results in \eqref{obsExp}.

\end{document}